\documentclass[runningheads]{llncs}
\usepackage{graphicx}

\usepackage[hidelinks]{hyperref}

\usepackage{color}
\usepackage{amsmath,amssymb}
\usepackage{nth}
\usepackage{siunitx}
\usepackage{microtype}

\usepackage[nameinlink]{cleveref}

\Crefname{figure}{Fig.}{Figs.}

\usepackage{xspace}
\makeatletter
\DeclareRobustCommand\onedot{\futurelet\@let@token\@onedot}
\def\@onedot{\ifx\@let@token.\else.\null\fi\xspace}

\def\ie{\emph{i.e}\onedot}

\makeatother

\begin{document}
\title{
Automatic Probe Movement Guidance for Freehand Obstetric Ultrasound
}
%
%
\author{
Richard Droste\inst{1} \and
Lior Drukker\inst{2} \and
Aris T.\ Papageorghiou\inst{2} \and
J.\ Alison Noble\inst{1}
}
%
\authorrunning{
R.\ Droste et al.
}
%
\institute{
Institute of Biomedical Engineering,
\mbox{University of Oxford}, Oxford, UK\\
\email{richard.droste@eng.ox.ac.uk}
\and
Nuffield Department of Women\textsc{\char13}s \& Reproductive Health, \mbox{University of Oxford, Oxford, UK}
}
\maketitle              
\begin{abstract}
We present the first system that provides real-time probe movement guidance for acquiring standard planes in routine freehand obstetric ultrasound scanning.
Such a system can contribute to the worldwide deployment of obstetric ultrasound scanning  by lowering the required level of operator expertise.
The system employs an artificial neural network that receives the ultrasound video signal and the motion signal of an inertial measurement unit (IMU) that is attached to the probe, and predicts a guidance signal.
The network termed US-GuideNet predicts either the movement towards the standard plane position (goal prediction), or the next movement that an expert sonographer would perform (action prediction).
While existing models for other ultrasound applications are trained with simulations or phantoms, we train our model with real-world ultrasound video and probe motion data from 464 routine clinical scans by 17 accredited sonographers.
Evaluations for 3 standard plane types show that the model provides a useful guidance signal with an accuracy of \SI{88.8}{\percent} for goal prediction and \SI{90.9}{\percent} for action prediction.
\end{abstract}
\section{Introduction}
Ultrasound scanning is an indispensable diagnostic tool in obstetrics due to its safety, real-time results and low cost.
At the same time, many women in developing countries do not receive a single ultrasound examination throughout their pregnancy due to a lack of skilled operators~\cite{Shah2015}.
The main tasks of ultrasound scanning are the acquisition, examination/verification and interpretation of pre-defined standard anatomical planes that enable the detection of fetal abnormalities.
Systems that provide assistance for or automate these tasks have the potential of enabling worldwide access to ultrasound scanning by reducing the level of necessary expertise.
Standard plane examination/verification and interpretation are largely standardized~\cite{Salomon2011}, can be performed remotely~\cite{Britton2019}, and can be facilitated through automated image analysis~\cite{Yaqub2017a}.
Freehand standard plane acquisition, on the other hand, is harder to facilitate/automate since it is not standardized and requires interaction with the mother.
It demands years of training and is the rate-limiting step even for experienced sonographers~\cite{Bahner2016}.

\begin{figure}[t]
{
\centering
\scriptsize
\sffamily
\def\svgwidth{1.\textwidth}
\graphicspath{{fig/}}
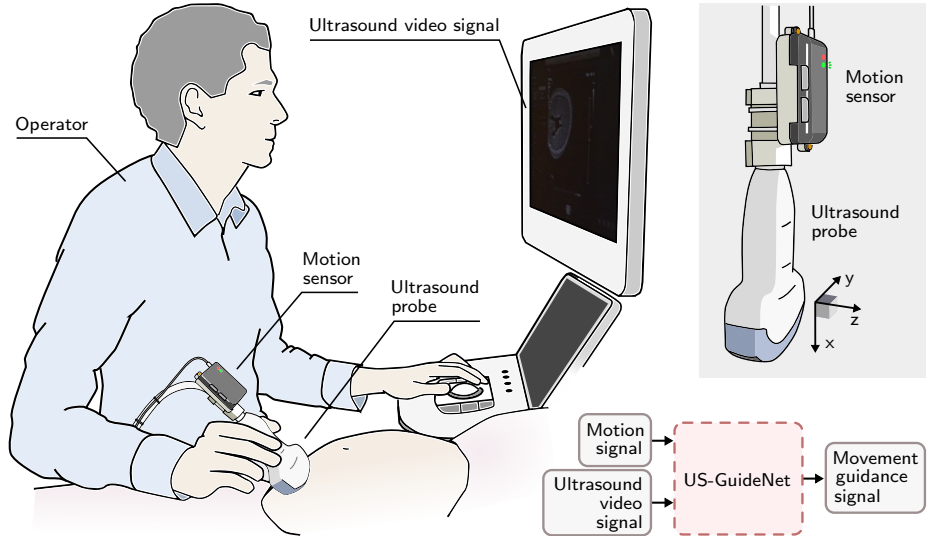
}
\caption{
System overview.
\emph{Left}: An operator performs ultrasound scanning with a routine clinical setup while the motion of the probe is recorded with an IMU.
\emph{Bottom right}: The \emph{US-GuideNet} receives the IMU motion signal and the ultrasound video signal as inputs and outputs a real-time probe movement guidance signal.
\emph{Top right}: Attachment of the IMU to the ultrasound probe and IMU coordinate system.
}
\label{fig:setup}
\end{figure}

To address this issue, we present the first system that provides real-time probe movement guidance for fetal standard plane acquisition in routine freehand obstetric ultrasound scanning. An overview of the system is presented in \Cref{fig:setup}.
An artificial neural network termed \emph{US-GuideNet} receives the ultrasound video signal alongside the signal of a motion sensor that is attached to the ultrasound probe, and outputs probe movement guidance that directs the operator towards the desired standard plane.
No specialized equipment is required: The motion sensor is a common inertial measurement unit (IMU) that is attached to the probe of a standard clinical ultrasound machine.
Further, the \emph{US-GuideNet} neural network is designed to be extremely lightweight and can run real time inference on a CPU.
Behavioral cloning (BC), a type of imitation learning, has emerged as a powerful technique to train neural networks to perform complex real-world tasks such as autonomous driving \cite{Pan2019}.
Here, we collect 5079 demonstrations of standard plane acquisitions from 464 \nth{2}- and \nth{3}-trimester scans acquired by 17 accredited sonographers and implement two settings of BC:
1) For \emph{goal prediction}, the network predicts the movement that leads directly to the estimated position of the standard plane.
2) For \emph{action prediction}, the network predicts the next movement that the expert would perform.

\section{Related Work}
Various approaches have been proposed to address the difficulty of ultrasound standard plane acquisition.

\textbf{Robotic Ultrasound.}
Human-controlled robotic systems have been developed that allow experienced sonographers to perform obstetric ultrasound exams remotely \cite{Vilchis2003}.
Automated robotic systems have been proposed for highly structured tasks such as finding planes of motionless objects \cite{Mebarki2010,Liang2010} or the human liver \cite{Mustafa2013}.
However, despite on-going efforts \cite{Wang2019c}, no robotic system has been proposed that can automate the complex task of obstetric ultrasound scanning.

\textbf{Simplified Acquisition Protocols.}
Instead of assisting operators to acquire typical freehand 2D standard planes, previous work has proposed to automatically extract standard planes from data that are acquired with a simplified protocol, such as 3D ultrasound volumes \cite{Rahmatullah2011a,Li2018d} or linear sweeps over the maternal abdomen \cite{Maraci2017}.
Moreover, IMUs have been used to acquire 3D ultrasound volumes with 2D probes \cite{Housden2008,Prevost2018}.
However, these methods are applicable only for a subset of standard planes (fetal abdomen and head) and the standard plane quality is not up to par with typical freehand scanning.

\textbf{Phantoms and Simulated Environments.}
Recent studies have proposed learning based systems that are trained to acquire ultrasound planes in simplified environments.
One study proposes an algorithm that learns to find a view of the adult heart in a grid of pre-acquired ultrasound images \cite{Milletari2019a}.
Moreover, learning-based systems have been proposed in which a robotic actuator finds predefined views of simple tissue phantoms \cite{Jarosik2019} or a fetal US phantom \cite{Toporek2018a}.
However, a fetus in the mother's womb is a dynamic and highly variant object that can not be well-represented with static simulations or a phantom.
Furthermore, these algorithms are purely image-based and therefore rely on the exact execution of the predicted actions, which is only possible within a simulation or with a robotic system.
Here, we train a guidance algorithm with video and probe motion data from a large number of real-world expert demonstrations from routine scanning.
Moreover, our algorithm receives the real-time probe motion signal and can therefore react to the movements of a human operator.

\begin{figure}[t]
{
\centering
\scriptsize
\sffamily
\def\svgwidth{1.\textwidth}
\graphicspath{{fig/}}
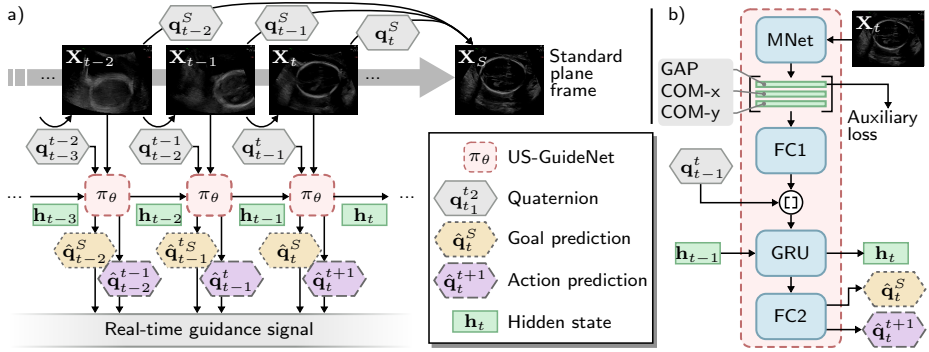
}
\caption{
a) Proposed behavioral cloning framework.
b) \emph{US-GuideNet} architecture.
}
\label{fig:method}
\end{figure}

\section{Method}
\label{sec:method}
\Cref{fig:setup} presents an overview of the proposed system.
An operator performs routine obstetric ultrasound scanning with a standard clinical machine.
An inertial measurement unit (IMU) motion sensor is attached to the ultrasound probe and an on-board attitude and heading reference system (AHRS) estimates the sensor's orientation in the earth coordinate system.
The motion sensor signal and the machine video signal are input into a neural network, \emph{US-GuideNet}, that outputs a 3D rotation of the probe that guides the operator towards the standard plane.
The network training method is described in \Cref{sec:lfd}, the network architecture is detailed in \Cref{sec:arch} and implementation details are provided in \Cref{sec:deets}.

\subsection{Learning from Expert Demonstrations}
\label{sec:lfd}
We pose the problem of training a neural network to predict a probe guidance signal as a behavioral cloning problem.
That is, we record standard plane acquisition demonstrations from several experts for a large number of patients and train the network to replicate the demonstrated behavior.
In general, a standard plane acquisition consists of live B-mode scanning followed by \emph{freezing} the ultrasound video and optionally selecting a previous frame with the desired appearance from a \emph{cine-buffer}.
We define the finally selected frame as the standard plane.

\textbf{Problem Formulation.}
\Cref{fig:method}~a) presents the formulation of the learning problem.
Let $\lbrace\mathbf{X}_t \in \mathbb{R}^{H\times W} \vert\, t\in\mathcal{T} \rbrace$ be ultrasound video frames of a standard plane acquisition, temporally downsampled to \SI{6}{\Hz}, with resolution $H\times W$ and frame indices $\mathcal{T} = \lbrace i \rbrace_{i=0}^F$, where $F$ is the \emph{freeze} frame index.
Moreover, let $\mathbf{X}_S$ be the standard plane with index $S\in\mathcal{T}\setminus\lbrace 0 \rbrace$.
Finally, let $\mathbf{q}_t = [q_w, q_x, q_y, q_z]^\top$ be the probe orientation quaternion of frame $\mathbf{X}_t$ and $\mathbf{q}_{t}^{t_2} := \mathbf{q}_{t}^\ast \mathbf{q}_{t_2}$ the probe rotation quaternion from frame $\mathbf{X}_t$ to frame $\mathbf{X}_{t_2}$, where $\mathbf{q}^\ast$ is the conjugate.
We represent orientations with quaternions since they can be smoothly interpolated without discontinuities or singularities, and are numerically stable and computationally efficient~\cite{Pavllo2019}.
Euler angles, in contrast, another popular representation of rotations, suffer from discontinuities such as the \emph{gimbal lock}.
We do not consider probe translation in this work since the IMU is not suitable to estimate it accurately.

\textbf{Behavioral Cloning.}
We train a policy network $\pi_\theta: s_t \mapsto u_t$ termed \emph{US-GuideNet} with parameters $\theta$ that maps the state $s_t$ at time step $t\in\mathcal{T}$ to an action $u_t$.
We define the state as the tuple $s_t:=(\mathbf{X}_t, \mathbf{q}_{t-1}^t, \mathbf{h}_{t-1})$, where $\mathbf{h}_{t-1}$ is the hidden state of a recurrent neural network within $\pi_\theta$.
We explore the two different settings for the action $u_t$: \emph{goal prediction} and \emph{action prediction}.
For \emph{goal prediction}, the policy $\pi_\theta^{g}: s^t \mapsto \hat{\mathbf{q}}_t^S$ estimates the rotation from the current orientation to the orientation of the standard plane.
If the estimated standard plane orientation is accurate, this policy is optimal, \ie, it guides the operator directly to the standard plane.
However, it is not guaranteed that enough information has been seen at time $t$ for an accurate estimation of the standard plane orientation.
Therefore, we explore a second setting, \emph{action prediction}, where the policy $\pi_\theta^a: s_t \mapsto \hat{\mathbf{q}}_t^{t+1}$ estimates the next rotation that the operator would perform.
This policy aims to closely mimic the expert sonographer behavior.

\textbf{Loss Function.}
During training, a demonstration is constructed from a subset of indices $\mathcal{T}_D \subset \mathcal{T}$ with start and end indices $t_0$ and $T$.
Let $\hat{\mathbf{Q}} := [\hat{\mathbf{q}}_t^{t_2}]_{t=t_0}^{T}$ and $\mathbf{Q} := [\mathbf{q}_t^{t_2}]_{t=t_0}^T$ be the predicted and ground truth rotation quaternion sequences, with $t_2\in \lbrace S, t+1 \rbrace$ for \emph{goal prediction} and \emph{action prediction} respectively.
We add an auxiliary output after the MNet in order to facilitate and regularize its training:
Since we want the MNet to recognize the appearance of standard planes, we input the average pooled MNet features into a softmax layer that predicts the class probabilities of the SonoNet standard plane classifier \cite{Baumgartner2017} for each frame.
Let $\hat{\mathbf{P}} = [\hat{\mathbf{p}}_t]_{t=t_0}^{T}$ be the auxiliary softmax output and $\mathbf{Y} = [\mathbf{y}_t]_{t=t_0}^{T}$ the SonoNet class probabilities, with $\hat{\mathbf{p}}_t, \mathbf{y}_t \in\mathbb{R}_{\geq 0}^{14}$.
The total training loss $\mathcal{L}$ is
\begin{equation}\nonumber
\mathcal{L} = \sum_{t=t_0+W}^T \Big\lbrace
\underbrace{
- \frac{1}{\lVert \hat{\mathbf{q}}_t^{t_2} \rVert} \hat{\mathbf{q}}_t^{t_2} \cdot \mathbf{q}_t^{t_2}
}_\text{Similarity loss}
+
\alpha 
\underbrace{
\vphantom{\frac{1}{\lVert \hat{\mathbf{q}}_t^{t_2} \rVert}}
(1 - \lVert \hat{\mathbf{q}}_t^{t_2} \rVert^2)^2
\Big\rbrace
}_\text{Norm loss}
- \beta \sum_{t=t_0}^T
\underbrace{
\vphantom{\frac{1}{\lVert \hat{\mathbf{q}}_t^{t_2} \rVert}}
\mathbf{y}_t^\top \mathrm{diag}(\mathbf{w})\log(\mathbf{p}_t)
}_\text{Auxiliary loss}
\end{equation}
where $\cdot$ denotes the dot-product, $\alpha, \beta \in \mathbb{R}_{>0}$ are scalar weighting factors, $\mathbf{w} \in\mathbb{R}_{\geq 0}^{14}$ is a weight vector that balances the SonoNet class probabilities, and $W$ is a warm up time for the rotation prediction.

\subsection{US-GuideNet Architecture}
\label{sec:arch}
The \emph{US-GuideNet} policy network receives the ultrasound video and probe motion signals and outputs predicted expert probe rotations as described in \Cref{sec:lfd}.
We design the architecture for small time and space computational complexity (runtime and model size) such that it can run real-time inference on the CPU of an inexpensive computer.
The network architecture is illustrated in \Cref{fig:method}~b).
At each time step $t$, the ultrasound video frame $X_t$ is fed into a MobileNet V2 (MNet) convolutional neural network \cite{Sandler2018}, which consists of a cascade of lightweight depthwise-separable and pointwise convolutions.
We use MNet with a width-multiplier of 0.5, \ie, 50\% reduced number of channels.
Next, the dimensionality of the MNet output is reduced with a custom \emph{ConcatPool} operation that preserves both semantic and spatial information by concatenating global average pooled (GAP) features with the x and y coordinates of the centers of mass (COM-x/y) of the feature maps.
After reducing the features to 128 channels with a fully-connected layer \emph{FC1}, they are concatenated with the current probe rotation quaternion $\mathbf{q}_{t-1}^t$ and input into a gated recurrent unit \cite{Cho2014a} (GRU) with 132 input channels and 128 hidden channels.
Finally \emph{FC2}, a fully-connected layer with one 128-channel hidden layer, outputs the 4-dimensional probe rotation quaternion.

\subsection{Experimental Setup}
\label{sec:deets}
\textbf{Data Acquisition.}
The data were acquired as part of the PULSE (Perception Ultrasound by Learning Sonographic Experience) project, a prospective study of routine fetal ultrasound scans performed in all trimesters by accredited sonographers and fetal medicine doctors at the maternity ultrasound unit, Oxford University Hospitals NHS Foundation Trust, Oxfordshire, United Kingdom.
The exams were performed on a GE Voluson E8 scanner (General Electric, USA) while the video signal of the machine monitor was recorded lossless at 30 Hz.
The motion of each of two curved linear array transducer (2D) probes was recorded with a NGIMU IMU/AHRS (x-io Technologies Ltd., UK).
Each motion sensor was attached to the cable outlet of the probe with a custom 3D-printed mount as shown in \Cref{fig:setup}.
The probe orientation quaternions were sampled at \SI{400}{\Hz}.
This study was approved by the UK Research Ethics Committee (Reference 18/WS/0051) and written informed consent was given by all participating pregnant women and operators.
In this paper we use ultrasound video and corresponding gaze data of 464 second and third trimester scans acquired by 17 accredited sonographers between May 2018 and February 2020.

\textbf{Data Processing.}
We extracted the standard plane acquisitions from the ultrasound scans with a purpose-built program based on optical character recognition.
For each of the 5079 resulting acquisitions,
the program outputs the corresponding live B-Mode scanning segment, the \emph{freeze} frame and the \emph{cine-buffer}-corrected standard plane.
In addition, the program labels acquisitions of the biometry standard planes: the femur standard plane (FSP), the abdominal standard plane (ASP) and the trans-ventricular plane (TVP) \cite{Salomon2011}, which we use for evaluation.
The acquisition duration was limited to \SI{10}{\second} before the standard plane.
We automatically corrected any lag between the video and motion signals by correlating frame differences with probe motion, and manually verified the synchronization.
The video frames were cropped such that the ultrasound machine's graphical user interface was removed, and normalized to zero-mean and unit-variance.
The scans are divided into 80\% for training and 20\% for testing.

\textbf{Training.}
For each training epoch, a demonstration of 32 frames is randomly selected from each standard plane acquisition, which corresponds to a duration \SI{5.3}{\second} at \SI{6}{\Hz}.
It is ensured that $\min_t\lbrace\mathbf{q}_t\cdot \mathbf{q}_S\rbrace \geq 0.7$ for each demonstration.
The frames are augmented by randomly changing of the brightness, contrast and gamma by $\pm 10\%$ and randomly symmetrically cropping up to 20\% of the frame border.
The frames are then down-sampled to the network input resolution of $224\times288$.
The MNet is pre-trained via the auxiliary loss with a large number of ultrasound frames.
The entire \emph{US-GuideNet} neural network is then trained from the demonstrations for 20 epochs with the AdamW optimizer \cite{Loshchilov2019} with weight decay of $10^{-2}$ and initial learning rate of 0.001, which is decayed by a factor of 0.1 every 8 epochs.
The batch size is set to 8 and the warm up time for the rotation loss to \SI{1}{\second}.
After training with all demonstrations, the model is fine-tuned for each evaluation plane (FSP, ASP and TVP) separately for 16 epochs.

\textbf{Evaluation and Baseline.}
We evaluate the trained model on the full-length standard plane acquisitions (clipped to \SI{10}{\second} before the standard plane).
For each time step, we classify the predicted probe rotation as correct if and only if $\angle (\mathbf{q}_t \hat{\mathbf{q}}_t^{t_2}, \mathbf{q}_{t_2}) < \angle (\mathbf{q}_t, \mathbf{q}_{t_2}), t_2\in\lbrace S, t+1\rbrace$, \ie, if applying the predicted rotation reduces the angle to the target orientation.
As before, $t_2=S$ for \emph{goal prediction} and $t_2$ = $t+1$ for \emph{action prediction}.
As a baseline rotation prediction we use $\mathbf{q}_{t-1}^t$, \ie, continuing in the current direction of rotation at each time step.

\begin{figure}[ht]
\includegraphics[width=\textwidth]{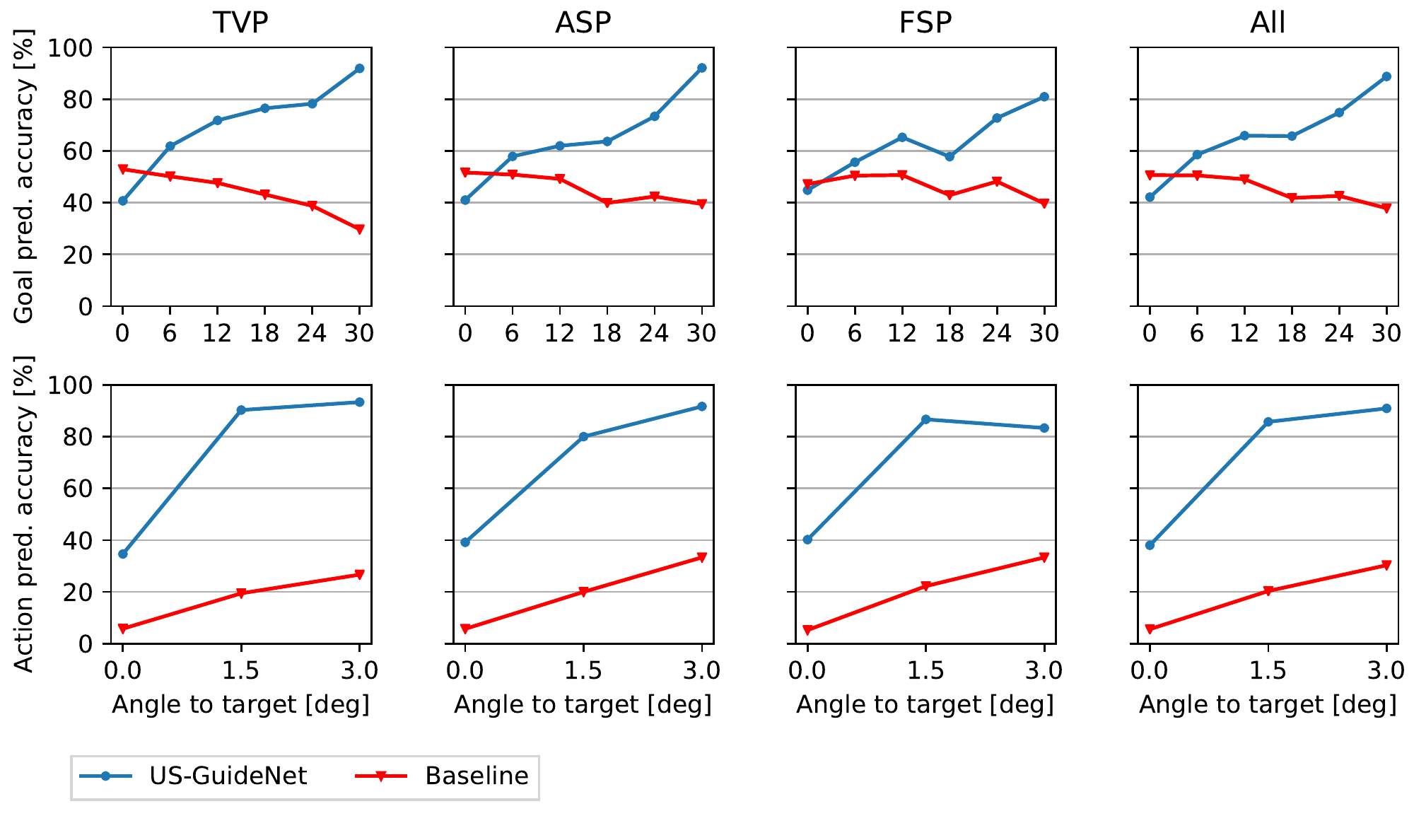}
\caption{
Experimental results for the evaluated standard planes: TVP (head), ASP (abdomen) and FSP (femur). In addition, the overall accuracies are provided.
}
\label{fig:results}
\end{figure}

\section{Results}
\label{sec:results}
The experimental results are shown in \Cref{fig:results}.
The average accuracy of the guidance signal and baseline is evaluated for different ranges of the angular distance to the target (standard plane orientation for \emph{goal prediction} or next probe position for \emph{action prediction}).
This enables the separation of the performance for coarse (large angular distance) and fine (low angular distance) adjustments.
The x-axis of the individual plots provides the lower limits of the ranges, which extend to the next-higher x-axis value.
Across the \emph{action prediction} and \emph{goal prediction} settings and all evaluated standard plane types, a common pattern can be observed that the accuracy of the guidance signal tends to increase with increasing angular distance to the standard plane.

\textbf{Goal Prediction.}
The \emph{goal prediction} accuracies are given in the upper row of \Cref{fig:results}.
The guidance signal performs better than the baseline for any angle range \SI{>6}{\degree}.
The accuracy of the guidance signal increases with increasing angular distances to the standard plane, ranging from \SI{42.2}{\percent} for angles \SIrange{0}{6}{\degree} to \SI{88.8}{\percent} for angles \SI{>30}{\degree}, with \SI{81.0}{\percent} for the FSP and \SI{92.1}{\percent} for the ASP.
The average baseline accuracy slightly declines towards higher angular distances.

\textbf{Action Prediction.}
The guidance signal accuracy is higher than the baseline accuracy for all target distance ranges.
The average guidance signal accuracy increases from \SI{38.0}{\percent} for angles \SIrange{0}{1.5}{\degree} to \SI{90.9}{\percent} for distances \SI{>3}{\degree}.
At angles \SI{>3}{\degree}, the largest accuracy is observed for the TVP with \SI{93.3}{\percent} and the lowest for the FSP with \SI{83.3}{\percent}.
The average baseline accuracy slightly increases with increasing angular distances.

\section{Discussion and Conclusion}
The results presented in \Cref{sec:results} demonstrate that the proposed probe guidance system for obstetric ultrasound scanning indeed provides a useful navigation signal towards the respective target, which is the standard plane orientation for \emph{goal prediction} and the next expert movement for \emph{action prediction}.
The accuracy of \emph{US-GuideNet} increases for larger differences to the target orientation, which shows that the algorithm is robust for guiding the operator towards the target orientation from distant starting points.
For small distances, it is difficult to predict an accurate guidance signal since the exact target orientation may be subject to inter- and intra-sonographer variations or sensor uncertainty.
The accuracy is similar for \emph{goal prediction} and \emph{action prediction} but slightly higher for action prediction at intermediate angles to the target, which can be explained by the fact that the action is always based on the previously seen frames, while the target position might be yet unknown.

The guidance signal accuracies are generally the highest for the abdominal and head standard planes (ASP and TVP).
The accuracy for the femur standard plane (FSP) is slightly lower, which can be explained by the fact the the femur is part of an extremity and therefore subject to more fetal movement, which can make its final position unpredictable.
Moreover, the FSP is defined via two anatomical landmarks---the distal and proximal ends of the femur---while the ASP and TVP are determined by the appearance of more anatomical structures \cite{Salomon2011}.
This might make it more difficult for the model to predict the FSP position that was chosen by the operator, since it is subject to more degrees of freedom.

A limitation of our study is that we test our algorithm with pre-acquired data.
However, in contrast with previous work \cite{Milletari2019a,Jarosik2019,Toporek2018a} which uses simulations or phantoms, our proposed system is trained and evaluated on data from real-world routine ultrasound scanning.
Moreover, instead of relying on the exact execution of the probe guidance as in previous work \cite{Milletari2019a,Jarosik2019,Toporek2018a}, our system reacts to the actual operator probe movements that are sensed with an IMU.
This suggests that the system will perform well in future tests on volunteer subjects.
In general, the accuracy of the predictions of \emph{US-GuideNet} is evident from the large improvements over the baseline of simply continuing the current direction of rotation.
While probe translation is not predicted due to IMU limitations, only the through-plane sweeping translation would usually change the view of the fetus while the sideways sliding and downwards/upwards translations would shift the fetal structure within the ultrasound image.
In combination with the rotation guidance, this leaves one degree of freedom to be determined by the operator.

In conclusion, this paper presents the first probe movement guidance system for the acquisition of standard planes in freehand obstetric ultrasound scanning.
Moreover, it is the first guidance system for any application of ultrasound standard plane acquisition that is trained with video and probe motion data from routine clinical scanning.
Our experiments have shown that the proposed \emph{US-GuideNet} network and behavioral cloning framework result in an accurate guidance system.
These results will serve as a foundation for subsequent validation studies with novice operators.
The proposed algorithm is lightweight which facilitates the deployment for existing ultrasound machines.

\begin{small}
\paragraph{Acknowledgements.}
We acknowledge the ERC (ERC-ADG-2015 694581, project PULSE) and the NIHR Oxford Biomedical Research Centre.
\end{small}


\end{document}